\documentclass{article}
\usepackage{amsfonts}

\newcommand{\be}{\begin{equation}}
\newcommand{\ee}{\end{equation}}
\newcommand{\bea}{\begin{eqnarray}}
\newcommand{\eea}{\end{eqnarray}}
\newcommand{\beaa}{\begin{eqnarray*}}
\newcommand{\eeaa}{\end{eqnarray*}}

\newcommand{\BB}{{{\rm I} \kern -2pt \rlap {\rm B} \kern +8pt}}

\advance\textheight by \topskip
\makeatletter

\@addtoreset{equation}{section}
\def\section{\@startsection {section}{1}{\z@}{-3.5ex plus -1ex minus
 -.2ex}{2.3ex plus .2ex}{\large\bf\centering}}
\def\subsection{\@startsection{subsection}{2}{\z@}{-3.25ex plus -1ex minus -.2ex}{1.5ex plus .2ex}{\bf}}
\def\subsubsection{\@startsection{subsubsection}{3}{\z@}{-3.25ex plus -1ex minus -.2ex}{1.5ex plus .2ex}{\sl}}
\makeatother

\begin{document}

\baselineskip 18pt \parindent 12pt \parskip 10pt

\begin{titlepage}

\begin{center}
{\Large {\bf The $U(N)$ chiral model and exact multi-solitons }}\\\vspace{1.5in} {\large Bushra Haider $^{a,}$  \footnote{%
bushrahaider@hotmail.com}  and M. Hassan $^{a,b,}$  \footnote{%
mhassan@physics.pu.edu.pk, mhassan@maths.gla.ac.uk }
}\vspace{0.15in}

{\small{\it $^{a}$Department of Physics,\\ University of the Punjab,\\
Quaid-e-Azam Campus,\\Lahore-54590, Pakistan.}}

{\small{\it $^{b}$Department of Mathematics,\\ University of Glasgow,\\
Glasgow G12 8QW, UK.}}
\end{center}

\vspace{1cm}
\begin{abstract}
We use binary Darboux transformation to obtain exact multi-soliton
solutions of principal chiral model and its noncommutative
generalization. We also show that the exact multi-solitons of
noncommutative principal chiral model in two dimensions and
noncommutative (anti-) self dual Yang-Mills equations in four
dimensions can be expressed explicitly in terms of
quasi-determinants.
\end{abstract}
\vspace{1cm} PACS: 11.10.Nx, 02.30.Ik
\end{titlepage}

\section{Introduction}It is well known that the classical principal
chiral model, a nonlinear sigma model with target manifold being a
Lie group, is an integrable field theoretic model in the sense
that it contains an infinite sequence of local and non-local
conserved quantities and can be embedded into the general scheme
of the inverse scattering method \cite{Brezin}-\cite{Evans1}$.$
Recently some investigations have been made regarding the
classical integrability of the principal chiral model with and
without a Wess-Zumino term and its
supersymmetric and noncommutative generalizations \cite{Evans1}-\cite{saleem}$%
.$ In these studies \cite{Evans1},\cite{Evans} the involution of
the local conserved quantities amongst themselves and with the
non-local conserved quantities has been investigated for the
bosonic models (with and without a Wess-Zumino term) and for their
supersymmetric generalization. It has been shown in \cite{saleem1}
that the supersymmetric generalization of the principal chiral
model admits a one-parameter family of superfield flat connections
that results in the existence of an infinite number of
conservation laws and a superfield Lax formalism.

The study of integrability of principal chiral model is also
important from the point of view of integrability and the
determination of the exact spectrum of free string theory on
$AdS^{5}\times S^{5}.$ In the last few years a great deal of work
has been done in studying the integrability of the classical
string sigma model on $AdS^{5}\times S^{5}$ (see e.g.
\cite{dorey}-\cite{hatsuda}). In these studies integrability
aspects such as the Lax formalism, the existence of local and
nonlocal conserved quantities, the fundamental Poisson bracket
algebra and the Yangian symmetry; has been investigated. It has
been shown that the infinite number of conserved quantities are in
involution for classical string moving on $\mathbf{R}\times S^{3}$
submanifold of $AdS^{5}\times S^{5} $. In fact, the bosonic
strings on $\mathbf{R}\times S^{3}$ are described in static gauge
by an $SU\left( 2\right) $ principal chiral model. From the point
of view of string theory, it seems natural to study the
construction of multi-soliton solutions for the principal chiral
model through a solution generating technique of binary Darboux
transformation. The purpose of this work is to use the binary
Darboux transformation for obtaining exact multi-soliton solutions
of the principal chiral model and the $N$-soliton solution has
been obtained in terms of ratio of determinants of certain
matrices.

The other aspect that we shall be interested in investigating is
the generalization of our results to the case of noncommutative
principal chiral model. In recent times, there has been an
increasing interest in the study of noncommutative integrable
field models (see e.g. \cite{saleem}-\cite{Dimakis1})$.$ The
interest in these studies is partly due to the fact that the
noncommutative field theories play an important role in string
theory,
D-brane dynamics, quantum Hall effect, etc. (see e.g. \cite{minwalla}-\cite%
{furuta} ). The noncommutative principal chiral model has been
studied
recently and it has been shown that the noncommutative generalization of $%
U\left( N\right) $ principal chiral model is integrable in the sense that it
contains an infinite sequence of conserved quantities and it admits a
one-parameter family of flat connections leading to a Lax formalism \cite%
{saleem}$.$ Moreover, it has been shown that the noncommutative
principal chiral model admits a solution generating technique of
elementary Darboux transformation which generates noncommutative
multi-soliton solutions of the model. In the present work, we
shall use the binary Darboux transformation to obtain the exact
noncommutative multi-soliton solution of the noncommutative
principal chiral model. We show that the exact multi-solitons of
noncommutative principal chiral model can be expressed in terms of
quasi-determinants introduced by Gelfand and Retakh
\cite{gr}-\cite{gr3}$.$ The quasi-determinants also appear in the
construction of soliton solutions of some integrable systems (see
e.g. \cite{hamanakag}-\cite{hassan}). We also compare our results
with those for the noncommutative (anti) self-dual Yang-Mills
theory that acts as a master theory in the sense of Ward
conjecture \cite{ward1}-\cite{ward3} which states that almost all
(noncommutative) integrable systems including the principal chiral
model can be obtained by (anti) self-dual Yang-Mills equations (or
its generalizations) by reduction \cite{hamanaka1},
\cite{hamanaka6}$.$

\section{The $U(N)$ principal chiral model}
The field variables $g\left( x\right) $ of the $U\left( N\right) $
principal chiral model (PCM) take values in the Lie group $U\left(
N\right) $ \cite{Brezin}-\cite{Evans1}. The
action for the $U\left( N\right) $ PCM can be written in terms of field $%
g\left( x\right) $ as \footnote{%
Our conventions are such that the two-dimensional coordinates are related as
$x^{\pm }=\frac{1}{2}\left( x^{0}\pm ix^{1}\right) $ and $\partial ^{\pm }=%
\frac{1}{2}\left( \partial _{0}\pm i\partial _{1}\right) .$}%
\begin{equation}
S=\frac{1}{2}\int d^{2}x\mbox{Tr}\left( \partial
_{+}g^{-1}\partial _{-}g\right) ,  \label{action}
\end{equation}%
with
\begin{equation}
g^{-1}\left( x^{+},x^{-}\right) g\left( x^{+},x^{-}\right) =g\left(
x^{+},x^{-}\right) g^{-1}\left( x^{+},x^{-}\right) =1,  \label{fields}
\end{equation}%
where $g\left( x^{+},x^{-}\right) \in U\left( N\right) .$ The $U\left(
N\right) $-valued field $g\left( x^{+},x^{-}\right) $ can be expressed as
\begin{equation}
g\left( x^{+},x^{-}\right) \equiv e^{i\pi _{a}T^{a}}=1+i\pi _{a}T^{a}+\frac{1%
}{2}\left( i\pi _{a}T^{a}\right) ^{2}+\cdots ,  \label{field exp}
\end{equation}%
where $\pi _{a}$ is in the Lie algebra $u\left( N\right) $ of the Lie group $%
U\left( N\right) $ and $T^{a},a=1,2,3,\ldots ,N^{2},$ are Hermitian matrices
with the normalization Tr$\left( T^{a}T^{b}\right) =-\delta ^{ab}$ and are
the generators of $U\left( N\right) $ in the fundamental representation
satisfying the algebra%
\begin{equation}
\left[ T^{a},T^{b}\right] =if^{abc}T^{c}, \label{generators}
\end{equation}%
where $f^{abc}$ are the structure constants of the Lie algebra $u\left(
N\right) $. For any $X\in u\left( N\right) $, we write $X=X^{a}T^{a}$ and $%
X^{a}=-$Tr$\left( T^{a}X\right) .$ The action (\ref{action}) is invariant
under a global continuous symmetry%
\begin{equation}
U_{L}\left( N\right) \times U_{R}\left( N\right) :\mbox{ \ \ \ \ \
\ \ \ \ \ \ }g\left( x^{+},x^{-}\right) \longmapsto ugv^{-1},
\label{gsymmetry}
\end{equation}%
where $u\in U_{L}\left( N\right) $ and $v\in U_{R}\left( N\right) .$ The
Noether conserved currents associated with the global symmetry of the PCM are%
\begin{equation}
j_{\pm }^{R}=-g^{-1}(\partial _{\pm }g),\mbox{ \ \ \ \ \ \ \ \ \ \
\ }j_{\pm }^{L}=(\partial _{\pm }g)g^{-1},  \label{lrcurrents}
\end{equation}%
which take values in the Lie algebra $u\left( N\right) ,$ so that one can
decompose the currents into components $j_{\pm }\left( x^{+},x^{-}\right)
=j_{\pm }^{a}\left( x^{+},x^{-}\right) T^{a}.$ The equation of motion
following from (\ref{action}) corresponds to the conservation of these
currents. The left and right currents satisfy the following conservation
equation%
\begin{equation}
\partial _{-}j_{+}+\partial _{+}j_{-}=0.  \label{continuity}
\end{equation}%
The currents also obey the zero-curvature condition%
\begin{equation}
\partial _{-}j_{+}-\partial _{+}j_{-}+\left[ j_{+},j_{-}\right] =0.
\label{z-c}
\end{equation}%
Equations (\ref{continuity}) and (\ref{z-c}) can also be expressed as
\begin{equation}
\partial _{-}j_{+}=-\partial _{+}j_{-}=-\frac{1}{2}\left[ j_{+},j_{-}\right]
.  \label{cz-c}
\end{equation}%
The equations (\ref{continuity})-(\ref{cz-c}) hold for both $j_{\pm }^{L}$
and $j_{\pm }^{R}.$

It is well known that the principal chiral model admits a
one-parameter family of flat currents \cite{Brezin}. We define a
one-parameter family of transformations
on the field $g\left( x^{+},x^{-}\right) $ as%
\begin{equation}
g\rightarrow g^{\left( \gamma \right) }=u^{\left( \gamma \right) }gv^{\left(
\gamma \right) -1},  \label{1-p}
\end{equation}%
where $\gamma $ is a parameter and $u^{\left( \gamma \right) },v^{\left(
\gamma \right) }$ are the matrices belonging to $U\left( N\right) $. We
choose the boundary values $u^{\left( 1\right) }=1,v^{\left( 1\right) }=1$
or $g^{\left( 1\right) }=g.$ The matrices $u^{\left( \gamma \right) }$and $%
v^{\left( \gamma \right) }$ satisfy the following set of linear equations%
\begin{eqnarray}
\partial _{\pm }u^{\left( \gamma \right) } &=&\frac{1}{2}\left( 1-\gamma
^{\mp 1}\right) j_{\pm }^{L}u^{\left( \gamma \right) },  \label{gamma1} \\
\partial _{\pm }v^{\left( \gamma \right) } &=&\frac{1}{2}\left( 1-\gamma
^{\mp 1}\right) j_{\pm }^{R}v^{\left( \gamma \right) }.  \label{gamma2}
\end{eqnarray}%
From now on, we shall consider right-hand currents and drop the superscript $%
R$ on the current to simply write $j_{\pm }^{R}=j_{\pm }.$ The compatibility
condition of the linear system (\ref{gamma2}) is given by%
\begin{equation}
\{(1-\gamma ^{-1})\partial _{-}j_{+}-(1-\gamma )\partial _{+}j_{-}+\left( 1-%
\frac{1}{2}\left( \gamma +\gamma ^{-1}\right) \left[ j_{+},j_{-}\right]
\right) \}v^{\left( \gamma \right) }=0.  \label{compatibility}
\end{equation}%
Under the one-parameter family of transformations, the Noether conserved
currents transform as%
\begin{equation}
j_{\pm }\mapsto j_{\pm }^{\left( \gamma \right) }=\gamma ^{\mp 1}v^{\left(
\gamma \right) -1}j_{\pm }v^{\left( \gamma \right) }.  \label{1-pcurrent}
\end{equation}%
The linear system (\ref{gamma2}) can also be expressed in the following well
known form
\begin{equation}
\partial _{\pm }v\left( x^{+},x^{-};\lambda \right) =A_{\pm }^{\left(
\lambda \right) }v\left( x^{+},x^{-};\lambda \right) ,  \label{linearsys}
\end{equation}%
where the fields $A_{\pm }^{\left( \lambda \right) }$ are given by
\begin{equation}
A_{\pm }^{\left( \lambda \right) }=\mp \frac{\lambda }{1\mp \lambda }j_{\pm
}.  \label{laxpair}
\end{equation}%
Here $\lambda $ is the spectral parameter and is related to $\gamma $ by $%
\lambda =\frac{1-\gamma }{1+\gamma }.$ The compatibility condition
of the
linear system (\ref{linearsys}) is the zero-curvature condition%
\begin{equation}
\left[ \partial _{+}-A_{+}^{\left( \lambda \right) },\partial
_{-}-A_{-}^{\left( \lambda \right) }\right] \equiv \partial
_{-}A_{+}^{\left( \lambda \right) }-\partial _{+}A_{-}^{\left( \lambda
\right) }+\left[ A_{+}^{\left( \lambda \right) },A_{-}^{\left( \lambda
\right) }\right] =0.  \label{z-c1}
\end{equation}%
In other words, we have defined a one-parameter family of connections $%
A_{\pm }^{\left( \lambda \right) }$ which are flat. The Lax operators
\begin{equation}
L_{\pm }^{\left( \lambda \right) }=\partial _{\pm }-A_{\pm }^{\left( \lambda
\right) },  \label{laxoperator}
\end{equation}%
obey the following equations%
\begin{equation}
\partial _{\mp }L_{\pm }^{\left( \lambda \right) }=\left[ A_{\mp }^{\left(
\lambda \right) },L_{\pm }^{\left( \lambda \right) }\right] .
\label{laxoequation}
\end{equation}%
This is the Lax equation and the set of operators $\left(
L,A\right) $ is the given Lax pair of the model
\cite{Brezin}-\cite{Novikov}$.$ The Lax formalism detailed above
can be used to generate an infinite number of local and non-local
conserved quantities and to construct multi-soliton solutions of
the model.

\section{Binary Darboux transformation and exact multi-solitons}%
The Lax pair of the PCM can be used to construct binary Darboux
transformation of the system \cite{Ustinov}. We proceed by writing
the Lax pair (direct Lax pair) as%
\begin{eqnarray}
\partial _{+}v &=&\frac{1}{1-\lambda }j_{+}v,  \nonumber \\
\partial _{-}v &=&\frac{1}{1+\lambda }j_{-}v,  \label{laxpair20}
\end{eqnarray}%
where we have used $\lambda \rightarrow \frac{1}{\lambda }.$ From (\ref%
{laxpair20}), we have%
\begin{eqnarray}
v^{-1}(\partial _{+}v)v^{-1} &=&\frac{1}{1-\lambda
}v^{-1}j_{+}vv^{-1},  \nonumber
\\
\partial _{+}v^{-1} &=&-\frac{1}{1-\lambda }v^{-1}j_{+}.  \label{dual1}
\end{eqnarray}%
Similarly%
\begin{equation}
\partial _{-}v^{-1}=-\frac{1}{1+\lambda }v^{-1}j_{-}.  \label{dual2}
\end{equation}%
Let us denote
\begin{equation}
v^{-1}\equiv \omega ,  \label{dualsoln}
\end{equation}%
then from equations (\ref{dual1}), (\ref{dual2}) and by analogy of direct
Lax pair, we define another Lax pair (dual Lax pair) for the matrix field $%
\omega $ with spectral parameter $\lambda ^{\prime }$ as%
\begin{eqnarray}
\partial _{+}\omega &=&-\frac{1}{1-\lambda ^{\prime }}\omega j_{+},
\nonumber \\
\partial _{-}\omega &=&-\frac{1}{1+\lambda ^{\prime }}\omega j_{-}.
\label{dualpair2}
\end{eqnarray}%
Now consider two solutions $v_{1}$ and $v_{2}$ of (\ref{laxpair20}), then%
\[
\partial _{+}\left( v_{1}^{-1}v_{2}\right) =(\partial
_{+}v_{1}^{-1})v_{2}+v_{1}^{-1}\partial _{+}\left( v_{2}\right) ,
\]%
using (\ref{laxpair20}) and (\ref{dual1}) in the above equation we get%
\begin{equation}
\partial _{+}\left( v_{1}^{-1}v_{2}\right) =0.  \label{soln}
\end{equation}%
Similarly%
\begin{equation}
\partial _{-}\left( v_{1}^{-1}v_{2}\right) =0.  \label{soln2}
\end{equation}%
From equations (\ref{soln}) and (\ref{soln2}), we have%
\[
v_{1}^{-1}v_{2}=C\left( \lambda \right) ,
\]%
or%
\[
v_{1}^{-1}=C\left( \lambda \right) v_{2}^{-1},
\]%
where $C\left( \lambda \right) $ is some arbitrary matrix function. Now
using (\ref{dualsoln}), we see that%
\begin{equation}
\omega \left( x,\lambda \right) =C\left( \lambda \right) v^{-1}\left(
x,\lambda \right) ,  \label{connection}
\end{equation}%
where $\omega \left( x,\lambda \right) $ and $v\left( x,\lambda \right) $
are the solutions of the dual and the direct Lax pairs respectively. The
matrix field $g\left( x\right) $ can be related to the solution $\omega $ of
the dual Lax pair by%
\begin{equation}
g\left( x\right) C\left( 0\right) =\omega \left( x,\lambda \right)
\mid _{\lambda =0}.  \label{initialsoln1}
\end{equation}%
Again from (\ref{laxpair20}), we see that at $\lambda =0$%
\begin{eqnarray*}
\partial _{+}v &=&j_{+}v, \\
(\partial _{+}v)v^{-1} &=&j_{+},
\end{eqnarray*}%
using (\ref{lrcurrents}) we have%
\begin{equation}
(\partial _{+}v)v^{-1}=(\partial _{+}g)g^{-1},  \label{initialsoln}
\end{equation}%
impying%
\begin{equation}
g\left( x\right) C\left( 0\right) =v\left( x,\lambda \right) \mid
_{\lambda =0}.  \label{initialsolna}
\end{equation}%
It follows from (\ref{laxpair20}) that the matrix function $v$ may
be chosen to satisfy the reality condition %
\begin{equation}
v^{\dagger }\left( \bar{\lambda}\right) =v^{-1}\left( \lambda
\right) .\label{unitarity}
\end{equation}%
Similar equation holds for $\omega .$

Let $\left\vert m\right\rangle $ be a column solution and $\left\langle
n\right\vert $ be a row solution of the Lax pairs (\ref{laxpair20}), (\ref%
{dualpair2}) with spectral parameters $\mu $ and $\nu $ respectively ($\mu
\neq \nu $). Through a projection operator $P$, the one-fold binary Darboux
transformation can be constructed to obtain new matrix solutions $v\left[ 1%
\right] $ and $\omega \left[ 1\right] $ satisfying the direct and dual Lax
pairs (\ref{laxpair20}) and (\ref{dualpair2}), respectively. The solutions $v%
\left[ 1\right] $ and $\omega \left[ 1\right] $ are related to the old
solutions $v$ and $\omega $ respectively by the following transformation%
\begin{eqnarray}
v\left[ 1\right] &=&\left( I-\frac{\mu -\nu }{\lambda -\nu }P\right) v,
\nonumber \\
\omega \left[ 1\right] &=&\omega \left( I-\frac{\mu -\nu }{\mu -\lambda
^{\prime }}P\right) ,  \label{bdt}
\end{eqnarray}%
For the reality condition (\ref{unitarity}) to be satisfied, we have%
\begin{eqnarray*}
\nu &=&(\bar{\mu})^{-1}, \\
P^{\dagger } &=&P=P^{2},
\end{eqnarray*}%
where the projector $P$ is defined as%
\begin{equation}
P=\frac{\left\vert m\right\rangle \left\langle n\right\vert
}{\left\langle n\mid m\right\rangle },  \label{projector}
\end{equation}%
with
\begin{equation}
\left\langle n\mid m\right\rangle
=\sum\limits_{i=1}^{N}n_{i}m_{i}. \label{mn}
\end{equation}%
The projector $P$ has been expressed in terms of the solutions of Lax pairs (%
\ref{laxpair20}) and (\ref{dualpair2}). Let $g$ be a known solution of the
PCM, the binary Darboux transformation gives a new solution $g\left[ 1\right]
$ given by
\begin{equation}
g\left[ 1\right] =\left( I+\frac{\mu -\nu }{\nu }P\right) g,  \label{g1dt}
\end{equation}%
where $v\mid _{\lambda =0}=g.$ The new solutions\ $v\left[ 1\right] $ and $%
\omega \left[ 1\right] $ satisfy the direct and dual Lax pairs (\ref%
{laxpair20}) and (\ref{dualpair2}) respectively, which shows the covariance
of the Lax pair of the PCM under the binary Darboux transformation, implying
that the conserved currents $j_{\pm }$ transform as%
\begin{eqnarray}
j_{+}\left[ 1\right] &=&j_{+}-\left( \mu -\nu \right) \partial _{+}P,
\nonumber \\
j_{-}\left[ 1\right] &=&j_{-}+\left( \mu -\nu \right) \partial _{-}P,
\label{trcurrents}
\end{eqnarray}%
Substituting equations (\ref{bdt}) and (\ref{trcurrents}) into the systems
we get%
\begin{eqnarray}
\partial _{+}v\left[ 1\right] &=&\frac{1}{1-\lambda }j_{+}\left[ 1\right] v%
\left[ 1\right] ,  \nonumber \\
\partial _{-}v\left[ 1\right] &=&\frac{1}{1+\lambda }j_{-}\left[ 1\right] v%
\left[ 1\right] ,  \label{directtr}
\end{eqnarray}%
and%
\begin{eqnarray}
\partial _{+}\omega \left[ 1\right] &=&-\frac{1}{1-\lambda ^{\prime }}\omega %
\left[ 1\right] j_{+}\left[ 1\right] ,  \nonumber \\
\partial _{-}\omega \left[ 1\right] &=&-\frac{1}{1+\lambda ^{\prime }}\omega %
\left[ 1\right] j_{-}\left[ 1\right] .  \label{dualtr}
\end{eqnarray}

The successive iterations of binary Darboux transformation produces the
transformed matrix solutions of direct and dual Lax pairs as%
\begin{eqnarray}
v\left[ K\right] &=&\left( I-\frac{\mu ^{\left( K\right) }-\nu ^{\left(
K\right) }}{\lambda -\nu ^{\left( K\right) }}P\left[ K\right] \right) \cdots
\left( I-\frac{\mu ^{\left( 1\right) }-\nu ^{\left( 1\right) }}{\lambda -\nu
^{\left( 1\right) }}P\left[ 1\right] \right) v,  \nonumber \\
\omega \left[ K\right] &=&\omega \left( I-\frac{\mu ^{\left( 1\right) }-\nu
^{\left( 1\right) }}{\mu ^{\left( 1\right) }-\lambda ^{\prime }}P\left[ 1%
\right] \right) \cdots \left( I-\frac{\mu ^{\left( K\right) }-\nu ^{\left(
K\right) }}{\mu ^{\left( K\right) }-\lambda ^{\prime }}P\left[ K\right]
\right) ,  \label{dtn}
\end{eqnarray}%
where%
\begin{equation}
P\left[ i\right] =\frac{\left\vert m^{\left( i\right) }\left[
i-1\right] \right\rangle \left\langle n^{\left( i\right) }\left[
i-1\right] \right\vert }{\left\langle n^{\left( i\right) }\left[
i-1\right] \mid m^{\left( i\right) }\left[ i-1\right]
\right\rangle },  \label{projectorn}
\end{equation}%
and $\left\vert m^{\left( i\right) }\left[ i-1\right] \right\rangle $ and $%
\left\langle n^{\left( i\right) }\left[ i-1\right] \right\vert
,(i=1,2,3,\ldots ,K)$ defined as%
\begin{eqnarray}
&&\left\vert m^{\left( i\right) }\left[ i-1\right] \right\rangle
\nonumber
\\
&=&\left( I-\frac{\mu ^{\left( i-1\right) }-\nu ^{\left(
i-1\right) }}{\mu ^{\left( i\right) }-\nu ^{\left( i-1\right)
}}P\left[ i-1\right] \right) \cdots \left( I-\frac{\mu ^{\left(
i\right) }-\nu ^{\left( i\right) }}{\mu ^{\left( i\right) }-\nu
^{\left( i\right) }}P\left[ i\right] \right)
\left\vert m^{\left( i\right) }\right\rangle ,  \nonumber \\
&&\left\langle n^{\left( i\right) }\left[ i-1\right] \right\vert
\nonumber
\\
&=&\left\langle n^{\left( i\right) }\right\vert \left( I-\frac{\mu
^{\left( i\right) }-\nu ^{\left( i\right) }}{\mu ^{\left( i\right)
}-\nu ^{\left( i\right) }}P\left[ i\right] \right) \cdots \left(
I-\frac{\mu ^{\left( i-1\right) }-\nu ^{\left( i-1\right) }}{\mu
^{\left( i-1\right) }-\nu ^{\left( i\right) }}P\left[ i-1\right]
\right) ,  \label{vectorsoln}
\end{eqnarray}%
are the matrix-column and matrix-row solutions of direct and dual
Lax pairs, with spectral parameters $\mu ^{\left( i \right) }$ and
$\nu ^{\left( i\right) }$ respectively.

If we write the general form of multi-soliton solution of direct Lax pair (%
\ref{laxpair20}) in terms of partial fraction as%
\begin{equation}
v\left[ K\right] =\left( I-\sum\limits_{j=1}^{K}\frac{R_{j}}{\lambda -\nu
^{\left( j\right) }}\right) v,  \label{partialsoln}
\end{equation}%
and use the fact that $v\left[ K\right] =0$ if $\lambda =\mu
^{\left(
i\right) },$ $v=\left\vert m^{\left( i\right) }\right\rangle $, we get the $%
K $th iteration formula in the form%
\begin{equation}
v\left[ K\right] =\left( I-\sum\limits_{i,j=1}^{K}\frac{\mu ^{\left(
j\right) }-\nu ^{\left( i\right) }}{\lambda -\nu ^{\left( j\right) }}\frac{%
\left\vert m^{\left( i\right) }\right\rangle \left\langle n^{\left( j\right)
}\right\vert }{\left\langle n^{\left( i\right) }\mid m^{\left( j\right)
}\right\rangle }\right) v.  \label{phin}
\end{equation}%
Similarly, by using $\omega \left[ K\right] =0$ for $\lambda
^{\prime }=\nu ^{\left( i\right) },\omega =\left\langle n^{\left(
i\right) }\right\vert ,$
we get%
\begin{equation}
\omega \left[ K\right] =\omega \left( I-\sum\limits_{i,j=1}^{K}\frac{\mu
^{\left( j\right) }-\nu ^{\left( i\right) }}{\mu ^{\left( i\right) }-\lambda
^{\prime }}\frac{\left\vert m^{\left( i\right) }\right\rangle \left\langle
n^{\left( j\right) }\right\vert }{\left\langle n^{\left( i\right) }\mid
m^{\left( j\right) }\right\rangle }\right) .  \label{chin}
\end{equation}%
The relation of $v\left[ K\right] $ with the new solution $g\left[
K\right] $
of equation (\ref{continuity}) gives%
\begin{equation}
g\left[ K\right] =\left( I+\sum\limits_{i,j=1}^{K}\frac{\mu ^{\left(
j\right) }-\nu ^{\left( i\right) }}{\nu ^{\left( j\right) }}\frac{\left\vert
m^{\left( i\right) }\right\rangle \left\langle n^{\left( j\right)
}\right\vert }{\left\langle n^{\left( i\right) }\mid m^{\left( j\right)
}\right\rangle }\right) g.  \label{ncurrent}
\end{equation}%
For expression (\ref{ncurrent}) to ensure the positive-definite solution of (%
\ref{continuity})
\begin{eqnarray}
\nu ^{\left( i\right) } &=&\left( \bar{\mu}^{\left( i\right) }\right) ^{-1},
\label{cond1} \\
\left\langle n^{\left( i\right) }\right\vert &=&\left( \left\vert m^{\left(
i\right) }\right\rangle \right) ^{\dagger }g^{-1}=\left\langle m^{\left(
i\right) }\right\vert g^{-1}.  \label{cond3}
\end{eqnarray}%
The solutions $v\left[ K\right] $ and $\omega \left[ K\right] $ expressed in
additive form (\ref{phin}) and (\ref{chin}) respectively, are subjected to
the reality condition (\ref{unitarity}). The condition (\ref{unitarity}) to
be satisfied, one requires that the projectrs $P\left[ i\right] $ be
Hermitian and mutually orthogonal, i.e.%
\begin{eqnarray}
P^{\dagger }\left[ i\right] &=&P\left[ i\right] =P^{2}\left[ i\right] ,
\nonumber \\
P\left[ i\right] P\left[ j\right] &=&0,\mbox{ \ \ \ \ \ \ \ \ \ \ \ for }%
i\neq j.  \label{proj}
\end{eqnarray}%
The solutions of the $U\left( N\right) $ principal chiral model
obtained here are same as the solutions obtained through the well
known dressing method of Zakharov and Shabat \cite{Zakharov},
\cite{Novikov}. In the dressing method, the solution of the system
is obtained by reducing the solution of the spectral problem to
that of a Riemann-Hilbert problem with zero. By using the
technique of complex analysis, the solution of the system is
expressed in terms of projectors that relate solutions of
Riemann-Hilbert problem in a simple algebraic form. In the binary
Darboux transformation method discussed here, we express the
solution of the system in terms of projectors that can be obtained
in terms of the solutions of the direct and dual Lax pairs of the
system. We combine two elementary Darboux transformations to
construct binary Darboux transformation that generates solutions
of the system in terms of given projectors. Let us now consider
the second iteration of binary Darboux transformation. Take $\mu
^{\left(
1\right) }=\mu ,\mu ^{\left( 2\right) }=\bar{\mu}^{-1}+\varepsilon ,$ and $%
\left\langle m^{\left( 1\right) }g^{-1}\mid m^{\left( 2\right)
}\right\rangle =O\left( \varepsilon \right) .$ Taking the limit $\varepsilon
\rightarrow 0,$ the coefficients of matrix $g\left[ 2\right] $ become%
\begin{equation}
g\left[ 2\right] _{ik}=\frac{\left\vert
\begin{array}{ccc}
g_{ik} & -\left\langle m^{\left( 1\right) }g^{-1}\mid g^{\left( k\right)
}\right\rangle & -\left\langle m^{\left( 2\right) }g^{-1}\mid g^{\left(
k\right) }\right\rangle \\
\ m_{i}^{(1)} & \frac{\left\langle m^{\left( 1\right) }g^{-1}\mid
m^{\left( 1\right) }\right\rangle }{\left\vert \mu \right\vert
^{2}-1} & \mu ^{-1}\left\langle A^{\dagger }g^{-1}\mid m^{\left(
1\right) }\right\rangle \\
\ m_{i}^{(2)} & \bar{\mu}^{-1}\left\langle m^{\left( 1\right)
}g^{-1}\mid A\right\rangle & \frac{\left\vert \mu \right\vert
^{2}}{1-\left\vert \mu \right\vert ^{2}}\left\langle m^{\left(
2\right) }g^{-1}\mid m^{\left( 2\right) }\right\rangle%
\end{array}%
\right\vert }{\left\vert
\begin{array}{cc}
\frac{\left\langle m^{\left( 1\right) }g^{-1}\mid m^{\left( 1\right)
}\right\rangle }{\left\vert \mu \right\vert ^{2}-1} & \mu ^{-1}\left\langle
A^{\dagger }g^{-1}\mid m^{\left( 1\right) }\right\rangle \\
\bar{\mu}^{-1}\left\langle m^{\left( 1\right) }g^{-1}\mid A\right\rangle &
\frac{\left\vert \mu \right\vert ^{2}}{1-\left\vert \mu \right\vert ^{2}}%
\left\langle m^{\left( 2\right) }g^{-1}\mid m^{\left( 2\right) }\right\rangle%
\end{array}%
\right\vert },  \label{matrixg}
\end{equation}%
where
\begin{eqnarray*}
A &=&\frac{\partial \left\vert m^{\left( 2\right) }\right\rangle }{\partial
\mu ^{\left( 2\right) }}\mid _{\mu ^{\left( 2\right) }=\bar{\mu}^{-1}}, \\
g &=&\left( g^{\left( 1\right) },\ldots ,g^{\left( N\right)
}\right), \\
\left\vert m^{\left( i\right) } \right\rangle &=& \left(
m_{i}^{\left( i\right) },\ldots ,m_{N}^{\left( i\right)
}\right)^{T},\mbox{ \ \ \ \ \ \ \ \ \ \ \ }%
i=1,2,\cdots,K.
\end{eqnarray*}%
Similarly applying binary Darboux transformation $K$ times, we see that the
coefficients of matrix $g\left[ K\right] $ are
\begin{eqnarray}
&&g\left[ K\right] _{ik}  \nonumber \\
&=&\frac{\left\vert
\begin{array}{ccccccc}
g_{ik} & -a_{1k} & -a_{2k} & -a_{3k} & -a_{4k} & \cdots & -a_{Kk}
\\
\ m_{i}^{(1)} & \frac{1}{\left\vert \mu \right\vert ^{2}-1}b_{11}
& \mu ^{-1}A_{11}^{\dagger } & \frac{1}{\left\vert \mu \right\vert
^{2}-1}A_{21}^{\dagger } & \mu ^{-1}A_{31}^{\dagger } &
\cdots & c_{1K}A_{(K-1)1}^{\dagger } \\
\ m_{i}^{(2)} & \bar{\mu}^{-1}A_{11} & \frac{%
\left\vert \mu \right\vert ^{2}}{1-\left\vert \mu \right\vert ^{2}}b_{22} &
\bar{\mu}^{-1}A_{31} & \frac{\left\vert \mu \right\vert ^{2}}{1-\left\vert
\mu \right\vert ^{2}}A_{41} & \cdots & c_{2K}A_{K1} \\
\ m_{i}^{(3)} & \frac{1}{\left\vert \mu \right\vert ^{2}-1}A_{12}
& \mu ^{-1}A_{22} & \frac{1}{\left\vert \mu
\right\vert ^{2}-1}b_{33} & \mu ^{-1}A_{42} & \cdots & c_{3K}A_{K2} \\
\ m_{i}^{(4)} & \bar{\mu}^{-1}A_{13} & \frac{%
\left\vert \mu \right\vert ^{2}}{1-\left\vert \mu \right\vert ^{2}}A_{23} &
\bar{\mu}^{-1}A_{33} & \frac{\left\vert \mu \right\vert ^{2}}{1-\left\vert
\mu \right\vert ^{2}}b_{44} & \cdots & c_{4K}A_{K3} \\
\vdots & \vdots & \vdots & \vdots & \vdots & \ddots & \vdots \\
\ m_{i}^{(K)} & c_{K1}A_{1(K-1)} & c_{K2}A_{2(K-1)} &
c_{K3}A_{3(K-1)} & c_{K4}A_{4(K-1)} & \cdots &
c_{kK}b_{KK}%
\end{array}%
\right\vert }{\left\vert
\begin{array}{cccccc}
\frac{1}{\left\vert \mu \right\vert ^{2}-1}b_{11} & \mu ^{-1}A_{11}^{\dagger
} & \frac{1}{\left\vert \mu \right\vert ^{2}-1}A_{21}^{\dagger } & \mu
^{-1}A_{31}^{\dagger } & \cdots & c_{1K}A_{(K-1)1}^{\dagger } \\
\bar{\mu}^{-1}A_{11} & \frac{\left\vert \mu \right\vert ^{2}}{1-\left\vert
\mu \right\vert ^{2}}b_{22} & \bar{\mu}^{-1}A_{31} & \frac{\left\vert \mu
\right\vert ^{2}}{1-\left\vert \mu \right\vert ^{2}}A_{41} & \cdots &
c_{2K}A_{K1} \\
\frac{1}{\left\vert \mu \right\vert ^{2}-1}A_{12} & \mu ^{-1}A_{22} & \frac{1%
}{\left\vert \mu \right\vert ^{2}-1}b_{33} & \mu ^{-1}A_{42} & \cdots &
c_{3K}A_{K2} \\
\bar{\mu}^{-1}A_{13} & \frac{\left\vert \mu \right\vert ^{2}}{1-\left\vert
\mu \right\vert ^{2}}A_{23} & \bar{\mu}^{-1}A_{33} & \frac{\left\vert \mu
\right\vert ^{2}}{1-\left\vert \mu \right\vert ^{2}}b_{44} & \cdots &
c_{4K}A_{K3} \\
\vdots & \vdots & \vdots & \vdots & \ddots & \vdots \\
c_{K1}A_{1(K-1)} & c_{K2}A_{2(K-1)} & c_{K3}A_{3(K-1)} & c_{K4}A_{4(K-1)} &
\cdots & c_{kK}b_{KK}%
\end{array}%
\right\vert },  \label{matrixgkf}
\end{eqnarray}%
where%
\begin{eqnarray*}
a_{ik} &=&\left\langle m^{\left( i\right) }g^{-1}\mid g^{\left( k\right)
}\right\rangle , \\
b_{ik} &=&\left\langle m^{\left( i\right) }g^{-1}\mid m^{\left( k\right)
}\right\rangle , \\
A_{ik} &=&\left\langle m^{\left( i\right) }g^{-1}\mid A_{k}\right\rangle ,
\\
A_{ik}^{\dagger } &=&\left\langle A_{i}^{\dagger }g^{-1}\mid m^{\left(
k\right) }\right\rangle .
\end{eqnarray*}%
Please note that we have used the following notation for the coefficients of
the entries in $K$-$th$ column of $g\left[ K\right] _{ik}$
\begin{equation}
c_{ij}=\left\{
\begin{array}{c}
\mu ^{-1}\mbox{ \ \ \ \ \ \ \ \ \ \ \  (for odd values of }i,j%
\mbox{),} \\
\bar{\mu}^{-1}\mbox{ \ \ \ \ \ \ \ \ \   (for even values
of }i,j\mbox{),} \\
\frac{1}{\left\vert \mu \right\vert ^{2}-1}\mbox{ \ \ \ \ \ \ \ \ \ \ \ \  (for }i%
\mbox{ odd, }j\mbox{ even),} \\
\frac{\left\vert \mu \right\vert ^{2}}{1-\left\vert \mu \right\vert ^{2}}%
\mbox{ \ \ \ \ \ \ \ \ \ \ \ \ (for }i\mbox{ even, }j\mbox{ odd),}%
\end{array}%
\right.  \label{coeffs}
\end{equation}%
while the entries in $K$-$th$ row of $g\left[ K\right] _{ik}$ are obtained
as
\begin{equation}
c_{ji}=\bar{c}_{ij}.  \label{coeffrow}
\end{equation}%
The solution $g\left[ K\right] _{ik}$ is the $K$-soliton solution of the $%
U\left( N\right) $ principal chiral model. It has been mentioned earlier
that such solutions can be constructed using the dressing method.

\section{Binary Darboux transformation for a noncommutative $U(N)$
principal chiral model}In this section, we study the binary
Darboux transformation of a noncommutative $U(N)$ principal chiral
model and obtain the multi-soliton solutions of the model in terms
of quasi-determinants. In recent years, a lot of investigations
have been made regarding the noncommutative generalization of
integrable models (see e.g. \cite{saleem}-\cite{Dimakis1}). In
reference \cite{Moriconi}, a noncommutative extension of
$U(N)$-principal chiral model (nc-PCM) has been presented and it
is concluded that this noncommutative extension gives no extra
constraints for the theory to be integrable. The non-local
conserved
quantities of nc-PCM have also been derived using the iterative method of Br%
\'{}%
ezin-Itzykson-Zinn-Justin-Zuber (BIZZ) \cite{Brezin1} and the Lax
formalism of the nc-PCM to derive conserved quantities has been
developed in \cite{saleem}.

One way of obtaining noncommutative field theories is by the replacement of
ordinary products of field functions in commutative field theories with
their star products($\star $-products) and the resulting theories are
realized as deformed theories from the commutative ones. The $\star $%
-product is defined for ordinary fields on flat spaces, explicitly
by \cite{moyal}%
\begin{eqnarray}
f\left( x\right) g\left( x\right) &\rightarrow &\left( f\star g\right)
\left( x\right) =\exp [\left( \frac{i}{2}\theta ^{\mu \nu }\partial _{\mu
}^{x_{1}}\partial _{\nu }^{x_{2}}\right) f\left( x_{1}\right) g\left(
x_{2}\right) \mid _{x_{1}=x_{2}=x}]  \nonumber \\
&=&f\left( x\right) g\left( x\right) +\frac{i\theta ^{\mu \nu }}{2}\partial
_{\mu }f\left( x\right) \partial _{\nu }g\left( x\right) +O\left( \theta
^{2}\right) ,  \label{strproduct}
\end{eqnarray}%
where $\partial _{\mu }^{x_{i}}=\frac{\partial }{\partial
x_{i}^{\mu }}.$ The noncommutativity of coordinates of the
Euclidean space $R^{D}$ is
defined as%
\begin{equation}
\left[ x^{\mu },x^{\nu }\right] =i\theta ^{\mu \nu },  \label{commutator}
\end{equation}%
where $\theta ^{\mu \nu }$ is the second rank antisymmetric real
constant tensor known as the deformation parameter. The $\star
$-product of functions carries intrinsically the noncommutativity
of the coordinates and is
associative i.e.%
\[
\left( f\star g\right) \star h=f\star \left( g\star h\right)
\]%
As can be seen from (\ref{strproduct}), the noncommutative field
theories reduce to the ordinary (commutative) field theories as
the deformation parameter $\theta $ goes to zero (for more details
see e.g. \cite{minwalla}-\cite{furuta} ) . Following section 2, we
define the action for the two-dimensional $U\left( N\right) $
noncommuative principal chiral model (nc-PCM) as \cite{saleem}%
\begin{equation}
S^{\star }=\frac{1}{2}\int d^{2}x\mbox{Tr}\left( \partial
_{+}g^{-1}\star
\partial _{-}g\right) ,  \label{nc-action}
\end{equation}%
with%
\begin{equation}
g^{-1}\left( x^{+},x^{-}\right) \star g\left( x^{+},x^{-}\right) =g\left(
x^{+},x^{-}\right) \star g^{-1}\left( x^{+},x^{-}\right) =1,
\label{nc-fields}
\end{equation}%
where $g\left( x^{+},x^{-}\right) \in U\left( N\right) .$ In this case, the $%
U\left( N\right) $-valued field $g\left( x^{+},x^{-}\right) $ is defined as
\begin{equation}
g\left( x^{+},x^{-}\right) \equiv e_{\star }^{i\pi _{a}T^{a}}=1+i\pi
_{a}T^{a}+\frac{1}{2}\left( i\pi _{a}T^{a}\right) _{\star }^{2}+\cdots ,
\label{nc-fieldexp}
\end{equation}%
The action (\ref{nc-action}) is invariant under a global continuous symmetry%
\begin{equation}
U_{L}\left( N\right) \times U_{R}\left( N\right) :\mbox{ \ \ \ \ \
\ \ \ \ \ \ }g\left( x^{+},x^{-}\right) \longmapsto u\star g\star
v^{-1}, \label{nc-gsymmetry}
\end{equation}%
and the corresponding Noether conserved currents of the nc-PCM are%
\begin{equation}
j_{\pm }^{\star R}=-g^{-1}\star (\partial _{\pm }g),\mbox{ \ \ \ \
\ \ \ \ \ \ \ }j_{\pm }^{\star L}=(\partial _{\pm }g)\star g^{-1},
\label{nc-lrcurrents}
\end{equation}%
which take values in the Lie algebra $u\left( N\right) $. The left and right
currents satisfy the following conservation equation%
\begin{equation}
\partial _{-}j_{+}^{\star }+\partial _{+}j_{-}^{\star }=0,
\label{nc-continuity}
\end{equation}%
and the zero-curvature condition%
\begin{equation}
\partial _{-}j_{+}^{\star }-\partial _{+}j_{-}^{\star }+\left[ j_{+}^{\star
},j_{-}^{\star }\right] _{\star }=0,  \label{nc-z-c}
\end{equation}%
where $\left[ ,\right] _{\star }$ is the commutator with respect to $\star $%
-product i.e. for any functions $f$ and $g$, $\left[ f,g\right] _{\star
}=f\star g-g\star f.$ It can be easily seen that the equations (\ref%
{nc-continuity}) and (\ref{nc-z-c}) appear as the compatibility condition of
the following set of linear equations (Lax pair)
\begin{equation}
\partial _{\pm }v\left( x^{+},x^{-};\lambda \right) =A_{\pm }^{\star \left(
\lambda \right) }\star v\left( x^{+},x^{-};\lambda \right) ,
\label{nc-linearsys}
\end{equation}%
where the fields $A_{\pm }^{\star \left( \lambda \right) }$ are given by
\begin{equation}
A_{\pm }^{\star \left( \lambda \right) }=\mp \frac{\lambda }{1\mp \lambda }%
j_{\pm }^{\star },  \label{nc-laxpair}
\end{equation}%
and $\lambda $ is the spectral parameter$.$

By analogy of section 3, the Lax pair of the nc-PCM can be used to construct
binary Darboux transformation of the system. We proceed by rewriting the Lax
pair (direct Lax pair) as%
\begin{eqnarray}
\partial _{+}v &=&\frac{1}{1-\lambda }j_{+}^{\star }\star v,
\label{nc-laxpair2} \\
\partial _{-}v &=&\frac{1}{1+\lambda }j_{-}^{\star }\star v.
\label{nc-laxpair20}
\end{eqnarray}%
The dual Lax pair for the matrix field $\omega $ with spectral parameter $%
\lambda ^{\prime }$ is given as%
\begin{eqnarray}
\partial _{+}\omega &=&-\frac{1}{1-\lambda ^{\prime }}\omega \star
j_{+}^{\star },  \label{nc-dualpair1} \\
\partial _{-}\omega &=&-\frac{1}{1+\lambda ^{\prime }}\omega \star
j_{-}^{\star }.  \label{nc-dualpair2}
\end{eqnarray}%
Since all the objects involved are of matrix nature, therefore, the binary
Darboux transformation can be constructed for the nc-PCM in the same way as
for the usual (commutative) PCM. Following the previous section, one arrives
at the transformation%
\begin{eqnarray}
v\left[ 1\right] &=&\left( I-\frac{\mu -\nu }{\lambda -\nu }P\right) \star v,
\nonumber \\
\omega \left[ 1\right] &=&\omega \star \left( I-\frac{\mu -\nu }{\mu
-\lambda ^{\prime }}P\right) ,  \nonumber \\
g^{\star }\left[ 1\right] &=&\left( I+\frac{\mu -\nu }{\nu }P\right) \star
g^{\star },  \label{ncbdt}
\end{eqnarray}%
where the $P$ is the projector, defined as%
\begin{equation}
P=\left\vert m\right\rangle \star (\left\langle m\mid n\right\rangle
)^{-1}\star \left\langle n\right\vert .  \label{nc-projector}
\end{equation}%
By applying the successive Darboux transformation on $v,\omega $ and $g$, we
arrive at the following solution%
\begin{eqnarray}
g^{\star }\left[ K\right] &=&[I+(\sum\limits_{i,j=1}^{K}\frac{\mu
^{\left( j\right)
}-\nu ^{\left( i\right) }}{\nu ^{\left( j\right) }}  \nonumber \\
&&\times \left\vert m^{\left( i\right) }\right\rangle \star \left(
\left\langle n^{\left( i\right) }\mid m^{\left( j\right) }\right\rangle
\right) ^{-1}\star \left\langle n^{\left( j\right) }\right\vert )]\star
g^{\star },  \label{ncnthit}
\end{eqnarray}%
where
\begin{eqnarray}
\nu ^{\left( i\right) } &=&\left( \bar{\mu}^{\left( i\right) }\right) ^{-1},
\label{nc-cond1} \\
\left\langle n^{\left( i\right) }\right\vert &=&\left( \left\vert
m^{\left( i\right) }\right\rangle \right) ^{\dagger }\star
(g^{\star})^{-1}=\left\langle m^{\left( i\right) }\right\vert
\star (g^{\star})^{-1},  \label{nc-cond3}
\end{eqnarray}%
The solution (\ref{ncnthit}) of nc-PCM is different from
(\ref{ncurrent}) of the usual PCM in the sense that the product of
functions has been replaced with the corresponding $\star
$-product. The difference will clearly show up when we take
explicit expressions of solution (\ref{ncnthit}) of nc-PCM. In the
usual (commutative) case the solution appears as a ratio of
determinants of certain functions but in the noncommutative case,
the solution appears as quasi-determinant. To show this we
consider the second iteration of the binary Darboux transformation. Take $%
\mu ^{\left( 1\right) }=\mu ,\mu ^{\left( 2\right) }=\bar{\mu}%
^{-1}+\varepsilon ,$ and $ \left\langle m^{\left( i\right)
}(g^{\star })^{-1}\mid m^{\left( j\right)}\right\rangle=O\left(
\varepsilon \right) .$ Taking the
limit $\varepsilon \rightarrow 0,$ the coefficients of matrix $g^{\star }%
\left[ 2\right] $ may then be written in terms of quasi-
determinant \footnote{%
 Let $A = (a_{ij})$ be a $N{\times}N$ matrix and $B = (b_{ij})$ be the inverse matrix of $A$, that is,
$A{\star}B = B{\star}A = 1$. Quasi-determinants of $A$ are defined
formally as the inverse of the elements of $B = A^{-1}: |A_{ij}| =
b^{-1}$. Quasi-determinants can be also given iteratively by:
\[
|A|_{ij}=a_{ij}-\sum a_{ip}{\star}|A_{ij}|_{pq}^{-1}{\star}a_{qj}.
\]%

The quasi-determinant for a $1\times 1$ matrix $A=(a_{11})$ is
\[
|A|_{11}=a_{11}
\]
For a $2\times 2$ matrix $A=\left(
\begin{array}{cc}
a_{11} & a_{12} \\
a_{21} & a_{22}%
\end{array}%
\right) ,$ there exist four quasi-determinants given as%
\begin{eqnarray*}
|A|_{11} &=&\left\vert
\begin{array}{cc}
\frame{\fbox{$a_{11}$}} & a_{12} \\
a_{21} & a_{22}%
\end{array}%
\right\vert =a_{11}-a_{12}\ \star a_{22}^{-1}\star \ a_{21}, \\
|A|_{21} &=&\left\vert
\begin{array}{cc}
a_{11} & a_{12} \\
\frame{\fbox{$a_{21}$}} & a_{22}%
\end{array}%
\right\vert =a_{21}-a_{22}\star \ a_{12}^{-1}\ \star a_{11}, \\
|A|_{12} &=&\left\vert
\begin{array}{cc}
a_{11} & \frame{\fbox{$a_{12}$}} \\
a_{21} & a_{22}%
\end{array}%
\right\vert =a_{12}-a_{11}\star \ a_{21}^{-1}\star \ a_{22}, \\
|A|_{22} &=&\left\vert
\begin{array}{cc}
a_{11} & a_{12} \\
a_{21} & \frame{\fbox{$a_{22}$}}%
\end{array}%
\right\vert =a_{22}-a_{21}\star \ a_{11}^{-1}\ \star a_{12.}
\end{eqnarray*}%

For more examples and properties of quasi determinants see%
\cite{gr}-\cite{gr3}.}
 as
\begin{equation}
g^{\star }\left[ 2\right] _{ik}=\left\vert
\begin{array}{ccc}
\frame{\fbox{$g_{ik}^{\star}$}} & -\left\langle m^{\left( 1\right)
}(g^{\star })^{-1}\mid g^{\star \left( k\right) }\right\rangle &
-\left\langle m^{\left(
2\right) }(g^{\star })^{-1}\mid g^{\star \left( k\right) }\right\rangle \\
\ m_{i}^{(1)} & \frac{\left\langle m^{\left( 1\right) }(g^{\star
})^{-1}\mid m^{\left( 1\right) }\right\rangle }{\left\vert \mu
\right\vert ^{2}-1} & \mu
^{-1}\left\langle A^{\dagger }(g^{\star })^{-1}\mid m^{\left( 1\right) }\right\rangle \\
\ m_{i}^{(2)} & \bar{\mu}^{-1}\left\langle m^{\left( 1\right)
}(g^{\star })^{-1}\mid A\right\rangle & \frac{\left\vert \mu
\right\vert ^{2}}{1-\left\vert \mu \right\vert ^{2}}\left\langle
m^{\left(
2\right) }(g^{\star })^{-1}\mid m^{\left( 2\right) }\right\rangle%
\end{array}%
\right\vert ,  \label{nc-matrixg}
\end{equation}%
where
\begin{eqnarray*}
A &=&\frac{\partial \left\vert m^{\left( 2\right) }\right\rangle }{\partial
\mu ^{\left( 2\right) }}\mid _{\mu ^{\left( 2\right) }=\bar{\mu}^{-1}}, \\
g^{\star } &=&\left( g^{\star \left( 1\right) },\ldots ,g^{\star \left(
N\right) }\right) .
\end{eqnarray*}%
The $K$-$th$ iteration of BDT leads to the coefficients of matrix $g\left[ K%
\right] $ as%
\begin{eqnarray}
&&g^{\star }\left[ K\right] _{ik}  \nonumber \\
&=&\left\vert
\begin{array}{ccccccc}
\frame{\fbox{$g_{ik}^{\star}$}} & a_{1k}^{\star } & a_{2k}^{\star
} &
a_{3k}^{\star } & a_{4k}^{\star } & \cdots & a_{Kk}^{\star } \\
\ m_{i}^{(1)} & \frac{1}{\left\vert \mu \right\vert
^{2}-1}b_{11}^{\star } & \mu ^{-1}A_{11}^{\star \dagger } &
\frac{1}{\left\vert \mu \right\vert ^{2}-1}A_{21}^{\star \dagger }
& \mu
^{-1}A_{31}^{\star \dagger } & \cdots & c_{1K}A_{(K-1)1}^{\star \dagger } \\
\ m_{i}^{(2)} & \bar{\mu}^{-1}A_{11}^{\star }
& \frac{\left\vert \mu \right\vert ^{2}}{1-\left\vert \mu \right\vert ^{2}}%
b_{22}^{\star } & \bar{\mu}^{-1}A_{31}^{\star } & \frac{\left\vert \mu
\right\vert ^{2}}{1-\left\vert \mu \right\vert ^{2}}A_{41}^{\star } & \cdots
& c_{2K}A_{K1}^{\star } \\
\ m_{i}^{(3)} & \frac{1}{\left\vert \mu
\right\vert ^{2}-1}A_{12}^{\star } & \mu ^{-1}A_{22}^{\star } & \frac{1}{%
\left\vert \mu \right\vert ^{2}-1}b_{33}^{\star } & \mu ^{-1}A_{42}^{\star }
& \cdots & c_{3K}A_{K2}^{\star } \\
\ m_{i}^{(4)} & \bar{\mu}^{-1}A_{13}^{\star }
& \frac{\left\vert \mu \right\vert ^{2}}{1-\left\vert \mu \right\vert ^{2}}%
A_{23}^{\star } & \bar{\mu}^{-1}A_{33}^{\star } & \frac{\left\vert \mu
\right\vert ^{2}}{1-\left\vert \mu \right\vert ^{2}}b_{44}^{\star } & \cdots
& c_{4K}A_{K3}^{\star } \\
\vdots & \vdots & \vdots & \vdots & \vdots & \ddots & \vdots \\
\ m_{i}^{(K)} & c_{K1}A_{1(K-1)}^{\star } &
c_{K2}A_{2(K-1)}^{\star } & c_{K3}A_{3(K-1)}^{\star } & c_{K4}A_{4(K-1)}^{%
\star } & \cdots & c_{KK}b_{KK}^{\star }%
\end{array}%
\right\vert ,  \label{matrixngk}
\end{eqnarray}%
where
\begin{eqnarray*}
a_{ik}^{\star } &=&-\left\langle m^{\left( i\right) }(g^{\star
})^{-1}\mid
g^{\star \left( k\right) }\right\rangle , \\
b_{ik}^{\star } &=&\left\langle m^{\left( i\right) }(g^{\star
})^{-1}\mid
m^{\left( k\right) }\right\rangle , \\
A_{ik}^{\star } &=&\left\langle m^{\left( i\right) }(g^{\star
})^{-1}\mid
A_{k}\right\rangle , \\
A_{ik}^{\star \dagger } &=&\left\langle A_{i}^{\dagger }(g^{\star
})^{-1}\mid
m^{\left( k\right) }\right\rangle , \\
\mu ^{\left( 1\right) } &=&\mu , \\
\mu ^{\left( i\right) } &=&\frac{1}{\bar{\mu}^{(i-1)}}+\varepsilon
,\mbox{ \
\ \ \ }i=2,3,\cdots ,(N-1), \\
\left\langle m^{\left( i\right) }(g^{\star })^{-1}\mid
m^{\left( j\right)}\right\rangle &=& O\left( \varepsilon \right) ,\mbox{ \ \ \ \ }i\neq j, \\
A_{i-1} &=&\frac{\partial \left\vert m^{\left( i\right) }\right\rangle }{%
\partial \mu ^{\left( i\right) }}\mid _{\mu ^{\left( i\right) }=\bar{\mu}%
^{-1}}.
\end{eqnarray*}%
The coefficients of the entries in $K$-$th$ row and column of $g^{\star }%
\left[ K\right] _{ik}$ are same as given in equations (\ref{coeffs}) and (%
\ref{coeffrow}). Note that in the commutative limit i.e. $\theta \rightarrow
0$, the quasi-determinants $\left( \ref{nc-matrixg}\right) $ and $\left( \ref%
{matrixngk}\right) $ reduce to the ratio of determinants $\left( \ref%
{matrixg}\right) $ and $\left( \ref{matrixgkf}\right) $
respectively.

\section{Relation to binary Darboux transformation for
noncommutative (anti) self-dual Yang-Mills equations}(Anti)
self-dual Yang-Mills ((A)SDYM) theory is a well known example of
multi-dimensional integrable systems \cite{ward1}-\cite{ward3}.
The (A)SDYM equations also act as master equations of many
integrable equations in the sense of Ward conjecture
\cite{ward1}-\cite{ward3}. The noncommutative generalization of
(A)SDYM equations and its integrability aspects have been
investigated recently (e.g. \cite{hamanaka}-\cite{siddiq}). Following \cite%
{siddiq}, we write the (A)SDYM equations on a four-dimensional
noncommutative
space\footnote{%
The coordinates $x_{\mu },\mu =0,1,2,3$ on $4$-dimensional noncommutative
Euclidean space $E^{4}$ are related to the coordinates on noncommutative
complex Euclidean space as
\begin{eqnarray*}
y &=&x_{0}+ix_{3},\mbox{ \ \ \ \ }\bar{y}=x_{0}-ix_{3}, \\
z &=&x_{1}+ix_{2},\mbox{ \ \ \ \ }\bar{z}=x_{1}-ix_{2}.
\end{eqnarray*}%
The Yang-Mills fields are $N\times N$ matrix-valued $1$-forms representing $%
U\left( N\right) $ connections with components%
\begin{eqnarray*}
A_{y}^{\star } &=&g^{-1}\star \partial _{y}g,\mbox{ \ \ \ \ }A_{\bar{y}%
}^{\star }=\bar{g}^{-1}\star \partial _{\bar{y}}\bar{g}, \\
A_{z}^{\star } &=&g^{-1}\star \partial _{z}g,\mbox{ \ \ \ \ }A_{\bar{z}%
}^{\star }=\bar{g}^{-1}\star \partial _{\bar{z}}\bar{g},
\end{eqnarray*}%
where $g,\bar{g}$ and their inverses with respect to $\star $-product $%
g^{-1},\bar{g}^{-1}$ are functions of $y,\bar{y},z,\bar{z}$ and
are matrices belonging to $U\left( N\right) $.}%
\begin{eqnarray}
\partial _{\bar{y}}\mathcal{J}_{y}^{\star }+\partial _{\bar{z}}\mathcal{J}%
_{z}^{\star } &=&0,  \nonumber \\
\partial _{y}\mathcal{J}_{\bar{y}}^{\star }+\partial _{z}\mathcal{J}_{\bar{z}%
}^{\star } &=&0,  \label{asdym}
\end{eqnarray}%
where%
\begin{eqnarray*}
\mathcal{J}_{\bar{y}}^{\star } &=&\partial _{\bar{y}}J^{\star }\star
(J^{\star })^{-1}, \\
\mathcal{J}_{\bar{z}}^{\star } &=&\partial _{\bar{z}}J^{\star
}\star (J^{\star})^{-1}
\end{eqnarray*}%
and $ (J^{\star})^{-1} =\bar{g}^{-1}\star g$ is the inverse of $%
J^{\star }$ with respect to the $\star $-product. The noncommutative
(A)SDYM\ equations (nc-(A)SDYM equations) can also be expressed as the
compatibility condition of the following linear system (Lax pair)%
\begin{eqnarray}
(\partial _{y}+\lambda \partial _{\bar{z}})\Psi \left( y,\bar{y},z,\bar{z}%
;\lambda \right) &=&\mathcal{J}_{y}^{\star }\star \Psi \left( y,\bar{y},z,%
\bar{z};\lambda \right) ,  \nonumber \\
\left( \partial _{z}-\lambda \partial _{\bar{y}}\right) \Psi \left( y,\bar{y}%
,z,\bar{z};\lambda \right) &=&\mathcal{J}_{z}^{\star }\star \Psi \left( y,%
\bar{y},z,\bar{z};\lambda \right) ,\,  \label{asdym1}
\end{eqnarray}%
where $\Psi \left( y,\bar{y},z,\bar{z};\lambda \right) $ is some
$N\times N$ matrix-valued field and $\lambda $ is the spectral
parameter. The compatibility of the linear system (\ref{asdym1})
is
\[
(\partial _{z}\mathcal{J}_{y}^{\star }-\partial _{y}\mathcal{J}_{z}^{\star }+%
\left[ \mathcal{J}_{y}^{\star },\mathcal{J}_{z}^{\star }\right] )-\lambda
\left( \partial _{\bar{y}}\mathcal{J}_{y}^{\star }+\partial _{\bar{z}}%
\mathcal{J}_{z}^{\star }\right) =0.
\]%
Following \cite{siddiq}, it is easy to see that nc-(A)SDYM equations (\ref%
{asdym}) reduce to a noncommutative two-dimensional principal chiral field
equation, if we take $y=\bar{y}=x_{0}$ and $z=\bar{z}=x_{1}$,%
\[
\partial _{0}\mathcal{J}_{0}^{\star }+\partial _{1}\mathcal{J}_{1}^{\star
}=0,
\]%
where $\mathcal{J}_{0}^{\star }$ and $\mathcal{J}_{1}^{\star }$ are the
components of conserved currents associated with the global transformation $%
U\left( N\right) _{L}\times U\left( N\right) _{R}$.

The binary Darboux transformation of nc-(A)SDYM equation has been studied in
\cite{siddiq} where the $K$-soliton solution $J\left[ K\right] $ has been
expressed in terms of the projector $P\left[ i\right] $ as%
\[
J^{\star }\left[ K\right] =\left( I+\frac{\mu ^{\left( K\right) }-\nu
^{\left( K\right) }}{\nu ^{\left( K\right) }}P\left[ N\right] \right) \star
\cdots \star \left( I+\frac{\mu ^{\left( 1\right) }-\nu ^{\left( 1\right) }}{%
\nu ^{\left( 1\right) }}P\left[ 1\right] \right) \star J^{\star },
\]%
where%
\[
P\left[ i\right] =\left\langle \psi ^{\left( i\right) }\left[ i-1\right]
\right\vert \star \left\langle \phi ^{\left( i\right) }\left[ i-1\right]
\mid \psi ^{\left( i\right) }\left[ i-1\right] \right\rangle ^{-1}\star
\left\vert \phi ^{\left( i\right) }\left[ i-1\right] \right\rangle ,
\]%
and $\left\langle \psi ^{\left( i\right) }\left[ i-1\right] \right\vert $
and $\left\vert \phi ^{\left( i\right) }\left[ i-1\right] \right\rangle $
are row and column solutions of the direct and dual Lax pairs of nc-(A)SDYM
equations with spectral parameters $\mu ^{\left( i\right) }$ and $\nu
^{\left( i\right) }$ respectively. The $K$-soliton solution of nc-(A)SDYM
equations can also be expressed in terms of quasi-determinants as in the
case with nc-PCM. For this, we write (see \cite{Ustinov})
\[
J^{\star }\left[ 2\right] _{ik}=\left\vert
\begin{array}{ccc}
\frame{\fbox{$J_{ik}^{\star}$}} & -\left\langle \psi ^{\left(
1\right) }(J^{\star })^{-1}\mid J^{\star \left( k\right)
}\right\rangle & -\left\langle \psi ^{\left( 2\right) }(J^{\star
})^{-1}\mid J^{\star \left( k\right) }\right\rangle
\\
&  &  \\
\ \psi_{i}^{(1)} & \frac{\left\langle \psi ^{\left( 1\right)
}(J^{\star })^{-1}\mid \psi ^{\left( 1\right) }\right\rangle
}{\left\vert \mu \right\vert ^{2}-1} & \mu ^{-1}\left\langle
A^{\prime
\dagger }(J^{\star })^{-1}\mid \psi ^{\left( 1\right) }\right\rangle \\
&  &  \\
\ \psi_{i}^{(2)} & \bar{\mu}%
^{-1}\left\langle \psi ^{\left( 1\right) }(J^{\star })^{-1}\mid
A^{\prime }\right\rangle & \frac{\left\vert \mu \right\vert
^{2}}{1-\left\vert \mu \right\vert ^{2}}\left\langle \psi ^{\left(
2\right) }(J^{\star })^{-1}\mid
\psi ^{\left( 2\right) }\right\rangle \\
&  &
\end{array}%
\right\vert ,
\]%
where%
\begin{eqnarray}
\nu ^{\left( i\right) } &=&\left( \bar{\mu}^{\left( i\right) }\right) ^{-1},
\nonumber \\
\left\langle \phi ^{\left( i\right) }\right\vert &=&\left(
\left\vert \psi ^{\left( i\right) }\right\rangle \right) ^{\dagger
}\star (J^{\star })^{-1}=\left\langle \psi ^{\left( i\right)
}\right\vert \star (J^{\star })^{-1}, \label{cond}
\end{eqnarray}%
and%
\begin{eqnarray*}
A^{\prime } &=&\frac{\partial \left\vert \psi ^{\left( 2\right)
}\right\rangle }{\partial \mu ^{\left( 2\right) }}\mid _{\mu ^{\left(
2\right) }=\bar{\mu}^{-1}}, \\
J^{\star } &=&\left( J^{\star \left( 1\right) },\ldots ,J^{\star
\left( N\right) }\right),\\
\left\vert \psi^{\left( i\right) } \right\rangle &=& \left(
\psi_{1}^{\left( i\right) },\ldots ,\psi_{N}^{\left( i\right)
}\right)^{T},\mbox{ \ \ \ \ \ \ \ \ \ \ \ }%
i=1,2,\cdots,K .
\end{eqnarray*}%
\pagebreak
The $J^{\star }\left[ K\right] $ is now given as%
\begin{eqnarray*}
&&J^{\star }\left[ K\right] _{ik} \\
&=&\left\vert
\begin{array}{ccccccc}
\frame{\fbox{$J_{ik}^{\star}$}} & a_{1k}^{\prime \star } &
a_{2k}^{\prime \star } & a_{3k}^{\prime \star } & a_{4k}^{\prime
\star } & \cdots & a_{Kk}^{\prime
\star } \\
&  &  &  &  &  &  \\
\ \psi_{i}^{(1)} & \frac{1}{\left\vert \mu \right\vert
^{2}-1}b_{11}^{\prime \star } & \mu ^{-1}A_{11}^{\prime \star
\dagger } & \frac{1}{\left\vert \mu \right\vert
^{2}-1}A_{21}^{\prime \star \dagger } & \mu ^{-1}A_{31}^{\prime
\star \dagger } & \cdots &
c_{1K}A_{(K-1)1}^{\prime \star \dagger } \\
&  &  &  &  &  &  \\
\ \psi_{i}^{(2)} & \bar{\mu}%
^{-1}A_{11}^{\prime \star } & \frac{\left\vert \mu \right\vert ^{2}}{%
1-\left\vert \mu \right\vert ^{2}}b_{22}^{\prime \star } & \bar{\mu}%
^{-1}A_{31}^{\prime \star } & \frac{\left\vert \mu \right\vert ^{2}}{%
1-\left\vert \mu \right\vert ^{2}}A_{41}^{\prime \star } & \cdots &
c_{2K}A_{K1}^{\prime \star } \\
&  &  &  &  &  &  \\
\ \psi_{i}^{(3)} & \frac{1}{\left\vert \mu \right\vert
^{2}-1}A_{12}^{\prime \star } & \mu ^{-1}A_{22}^{\prime \star } &
\frac{1}{\left\vert \mu \right\vert ^{2}-1}b_{33}^{\prime \star }
& \mu
^{-1}A_{42}^{\prime \star } & \cdots & c_{3K}A_{K2}^{\prime \star } \\
&  &  &  &  &  &  \\
\ \psi_{i}^{(4)} & \bar{\mu}%
^{-1}A_{13}^{\prime \star } & \frac{\left\vert \mu \right\vert ^{2}}{%
1-\left\vert \mu \right\vert ^{2}}A_{23}^{\prime \star } & \bar{\mu}%
^{-1}A_{33}^{\prime \star } & \frac{\left\vert \mu \right\vert ^{2}}{%
1-\left\vert \mu \right\vert ^{2}}b_{44}^{\prime \star } & \cdots &
c_{4K}A_{K3}^{\prime \star } \\
&  &  &  &  &  &  \\
\vdots & \vdots & \vdots & \vdots & \vdots & \ddots & \vdots \\
\ \psi_{i}^{(K)} & c_{K1}A_{1(K-1)}^{\prime \star } &
c_{K2}A_{2(K-1)}^{\prime \star } & c_{K3}A_{3(K-1)}^{\prime \star
} & c_{K4}A_{4(K-1)}^{\prime \star } & \cdots &
c_{KK}b_{KK}^{\prime \star }
\\
&  &  &  &  &  &
\end{array}%
\right\vert ,
\end{eqnarray*}%
where%
\begin{eqnarray*}
a_{ik}^{\prime \star } &=&-\left\langle \psi ^{\left( i\right)
}(J^{\star })^{-1}\mid J^{\star \left( k\right) }\right\rangle , \\
b_{ik}^{\prime \star } &=&\left\langle \psi ^{\left( i\right) }(J^{\star })^{-1}\mid \psi ^{\left( k\right) }\right\rangle , \\
A_{ik}^{\prime \star } &=&\left\langle \psi ^{\left( i\right) }(J^{\star })^{-1}\mid A_{k}^{\prime }\right\rangle , \\
A_{ik}^{\prime \star \dagger } &=&\left\langle A_{i}^{\prime \dagger
}(J^{\star })^{-1}\mid \psi ^{\left( k\right) }\right\rangle , \\
\mu ^{\left( 1\right) } &=&\mu , \\
\mu ^{\left( i\right) } &=&\frac{1}{\bar{\mu}^{(i-1)}}+\varepsilon
,\mbox{ \
\ \ \ }i=2,3,\cdots ,(K-1), \\
\left\langle\psi ^{\left( i\right) }(J^{\star })^{-1}\mid \psi
^{\left( j\right)
}\right\rangle &=&O\left( \varepsilon \right) ,\mbox{ \ \ \ \ }i\neq j, \\
A_{i-1}^{\prime } &=&\frac{\partial \left\vert \psi ^{\left( i\right)
}\right\rangle }{\partial \mu ^{\left( i\right) }}\mid _{\mu ^{\left(
i\right) }=\bar{\mu}^{-1}}.
\end{eqnarray*}%
The coefficients of the entries in $K$-$th$ row and column of $J^{\star }%
\left[ K\right] _{ik}$ are same as given in equations (\ref{coeffs}) and (%
\ref{coeffrow}). Here again, one can see that in the commutative limit i.e. $%
\theta \rightarrow 0$, the quasi-determinants of multi-solitons reduce to
the ratio of determinants of multi-solitons of usual (commutative) (A)SDYM
equations. So in the limit $\theta \rightarrow 0$, we get the following
two-soliton solution as%
\[
J^{\star }\left[ 2\right] _{ik}\rightarrow J\left[ 2\right] _{ik}=\frac{%
\left\vert
\begin{array}{ccc}
J_{ik} & -\left\langle \psi ^{\left( 1\right) }J^{-1}\mid
J^{\left( k\right) }\right\rangle & \left\langle \psi ^{\left(
2\right)
}J^{-1}\mid J^{\left( k\right) }\right\rangle \\
&  &  \\
\ \psi_{i}^{(1)} & \frac{\left\langle \psi
^{\left( 1\right) }J^{-1}\mid \psi ^{\left( 1\right) }\right\rangle }{%
\left\vert \mu \right\vert ^{2}-1} & \mu ^{-1}\left\langle A^{\prime \dagger
}J^{-1}\mid \psi ^{\left( 1\right) }\right\rangle \\
&  &  \\
\ \psi_{i}^{(2)} & \bar{\mu}%
^{-1}\left\langle \psi ^{\left( 1\right) }J^{-1}\mid A^{\prime
}\right\rangle & \frac{\left\vert \mu \right\vert ^{2}}{1-\left\vert \mu
\right\vert ^{2}}\left\langle \psi ^{\left( 2\right) }J^{-1}\mid \psi
^{\left( 2\right) }\right\rangle \\
&  &
\end{array}%
\right\vert }{\left\vert
\begin{array}{cc}
\frac{\left\langle \psi ^{\left( 1\right) }J^{-1}\mid \psi ^{\left(
1\right) }\right\rangle }{\left\vert \mu \right\vert ^{2}-1} & \mu
^{-1}\left\langle A^{\prime \dagger }J^{-1}\mid \psi ^{\left( 1\right)
}\right\rangle \\
&  \\
\bar{\mu}^{-1}\left\langle \psi ^{\left( 1\right) }J^{-1}\mid A^{\prime
}\right\rangle & \frac{\left\vert \mu \right\vert ^{2}}{1-\left\vert \mu
\right\vert ^{2}}\left\langle \psi ^{\left( 2\right) }J^{-1}\mid \psi
^{\left( 2\right) }\right\rangle \\
&
\end{array}%
\right\vert },
\]
\pagebreak
and similarly for the multi-soliton solution, we get%
\begin{eqnarray*}
&&J^{\star }\left[ K\right] _{ik}\rightarrow J\left[ K\right] _{ik} \\
&=&\frac{\left\vert
\begin{array}{ccccccc}
J_{ik} & -a_{1k}^{\prime } & -a_{2k}^{\prime } & -a_{3k}^{\prime
}
& -a_{4k}^{\prime } & \cdots & -a_{Kk}^{\prime } \\
&  &  &  &  &  &  \\
\ \psi_{i}^{(1)} & \frac{1}{\left\vert \mu \right\vert
^{2}-1}b_{11}^{\prime } & \mu ^{-1}A_{11}^{\prime \dagger } &
\frac{1}{\left\vert \mu \right\vert ^{2}-1}A_{21}^{\prime \dagger
} & \mu ^{-1}A_{31}^{\prime \dagger } & \cdots &
c_{1K}A_{(K-1)1}^{\prime \dagger }
\\
&  &  &  &  &  &  \\
\ \psi_{i}^{(2)} & \bar{\mu}%
^{-1}A_{11}^{\prime } & \frac{\left\vert \mu \right\vert ^{2}}{1-\left\vert
\mu \right\vert ^{2}}b_{22}^{\prime } & \bar{\mu}^{-1}A_{31}^{\prime } &
\frac{\left\vert \mu \right\vert ^{2}}{1-\left\vert \mu \right\vert ^{2}}%
A_{41}^{\prime } & \cdots & c_{2K}A_{K1}^{\prime } \\
&  &  &  &  &  &  \\
\ \psi_{i}^{(3)} & \frac{1}{\left\vert \mu
\right\vert ^{2}-1}A_{12}^{\prime } & \mu ^{-1}A_{22}^{\prime } & \frac{1}{%
\left\vert \mu \right\vert ^{2}-1}b_{33}^{\prime } & \mu ^{-1}A_{42}^{\prime
} & \cdots & c_{3K}A_{K2}^{\prime } \\
&  &  &  &  &  &  \\
\ \psi_{i}^{(4)} & \bar{\mu}%
^{-1}A_{13}^{\prime } & \frac{\left\vert \mu \right\vert ^{2}}{1-\left\vert
\mu \right\vert ^{2}}A_{23}^{\prime } & \bar{\mu}^{-1}A_{33}^{\prime } &
\frac{\left\vert \mu \right\vert ^{2}}{1-\left\vert \mu \right\vert ^{2}}%
b_{44}^{\prime } & \cdots & c_{4K}A_{K3}^{\prime } \\
&  &  &  &  &  &  \\
\vdots & \vdots & \vdots & \vdots & \vdots & \ddots & \vdots \\
\ \psi_{i}^{(K)} & c_{K1}A_{1(K-1)}^{\prime } &
c_{K2}A_{2(K-1)}^{\prime } & c_{K3}A_{3(K-1)}^{\prime } &
c_{K4}A_{4(K-1)}^{\prime } & \cdots & c_{KK}b_{KK}^{\prime } \\
&  &  &  &  &  &
\end{array}%
\right\vert }{\left\vert
\begin{array}{cccccc}
\frac{1}{\left\vert \mu \right\vert ^{2}-1}b_{11}^{\prime } & \mu
^{-1}A_{11}^{\prime \dagger } & \frac{1}{\left\vert \mu \right\vert ^{2}-1}%
A_{21}^{\prime \dagger } & \mu ^{-1}A_{31}^{\prime \dagger } & \cdots &
c_{1K}A_{(K-1)1}^{\prime \dagger } \\
&  &  &  &  &  \\
\bar{\mu}^{-1}A_{11}^{\prime } & \frac{\left\vert \mu \right\vert ^{2}}{%
1-\left\vert \mu \right\vert ^{2}}b_{22}^{\prime } & \bar{\mu}%
^{-1}A_{31}^{\prime } & \frac{\left\vert \mu \right\vert ^{2}}{1-\left\vert
\mu \right\vert ^{2}}A_{41}^{\prime } & \cdots & c_{2K}A_{K1}^{\prime } \\
&  &  &  &  &  \\
\frac{1}{\left\vert \mu \right\vert ^{2}-1}A_{12}^{\prime } & \mu
^{-1}A_{22}^{\prime } & \frac{1}{\left\vert \mu \right\vert ^{2}-1}%
b_{33}^{\prime } & \mu ^{-1}A_{42}^{\prime } & \cdots & c_{3K}A_{K2}^{\prime
} \\
&  &  &  &  &  \\
\bar{\mu}^{-1}A_{13}^{\prime } & \frac{\left\vert \mu \right\vert ^{2}}{%
1-\left\vert \mu \right\vert ^{2}}A_{23}^{\prime } & \bar{\mu}%
^{-1}A_{33}^{\prime } & \frac{\left\vert \mu \right\vert ^{2}}{1-\left\vert
\mu \right\vert ^{2}}b_{44}^{\prime } & \cdots & c_{4K}A_{K3}^{\prime } \\
&  &  &  &  &  \\
\vdots & \vdots & \vdots & \vdots & \ddots & \vdots \\
c_{K1}A_{1(K-1)}^{\prime } & c_{K2}A_{2(K-1)}^{\prime } &
c_{K3}A_{3(K-1)}^{\prime } & c_{K4}A_{4(K-1)}^{\prime } & \cdots &
c_{KK}b_{KK}^{\prime } \\
&  &  &  &  &
\end{array}%
\right\vert },
\end{eqnarray*}%
where%
\begin{eqnarray*}
a_{ik}^{\prime } &=&\left\langle \psi ^{\left( i\right) }J^{-1}\mid
J^{\left( k\right) }\right\rangle , \\
b_{ik}^{\prime } &=&\left\langle \psi ^{\left( i\right) }J^{-1}\mid \psi
^{\left( k\right) }\right\rangle , \\
A_{ik}^{\prime } &=&\left\langle \psi ^{\left( i\right) }J^{-1}\mid
A_{k}^{\prime }\right\rangle , \\
A_{ik}^{\prime \dagger } &=&\left\langle A_{i}^{\prime \dagger }J^{-1}\mid
\psi ^{\left( k\right) }\right\rangle , \\
\mu ^{\left( 1\right) } &=&\mu , \\
\mu ^{\left( i\right) } &=&\frac{1}{\bar{\mu}^{(i-1)}}+\varepsilon
\qquad i=2,3,\cdots ,(K-1), \\
\left\langle \psi ^{\left( i\right) }J^{-1}\mid \psi ^{\left(
j\right)
}\right\rangle &=&O\left( \varepsilon \right) ,\qquad i\neq j, \\
A_{i-1}^{\prime } &=&\frac{\partial \left\vert \psi ^{\left(
i\right) }\right\rangle }{\partial \mu ^{\left( i\right) }}\mid
_{\mu ^{\left( i\right) }=\bar{\mu}^{-1}}.
\end{eqnarray*}%
The values of coefficients $c_{ij}$ are as given in equation (\ref{coeffs}).

\section{Conclusions} We have constructed a binary Darboux
transformation to generate exact multi-soliton solutions of
$U\left( N\right) $\ principal chiral model. The multi-soliton
solutions of the noncommutative $U\left( N\right) $ principal
chiral model are also obtained and the solutions are expressed in
terms of quasi-determinants of Gel'fand and Retakh. We find that
these solutions have the same form as that of (anti) self dual
Yang-Mills equations. Our results are useful in the sense that
their exact analysis leads to the various applications of D-brane
dynamics and helps understanding the properties of $N=2$ string
theory. This technique of binary Darboux transformation can also
be applied to other integrable models to obtain their exact
mulit-soliton solutions. These solutions may be analysed and it
would be interesting to check their stability and the scattering
properties. The work can be further extended to construct super
multi-solitons for the supersymmetric principal chiral model. From
the point of view of string theory, it is also interesting to
study the spectrum of solutions in the string theory on
$Ads^{5}\times S^{5}$ using the binary Darboux transformation.

\bigskip

{\large \textbf{Acknowledgements}}

BH would like to acknowledge the enabling role of the Higher
Education Commission, Pakistan and appreciates its financial
support through \textquotedblleft\ Indeginous 5000 fellowship
program" for PhD studies in Science and Technology. MH would like
to thank Jonathan Nimmo for hospitality at the Department of
Mathematics, University of Glasgow and to the Higher Education
Commission, Pakistan for a research fellowship.

\appendix

\bigskip

\end{document}